\documentclass[
aps,%
12pt,%
final,%
notitlepage,%
oneside,%
onecolumn,%
nobibnotes,%
nofootinbib,%
superscriptaddress,%
noshowpacs]%
{revtex4}
\usepackage{graphicx}
\usepackage{color}

\def\b{\begin{eqnarray}}
\def\e{\end{eqnarray}}

\def\f{\frac}
\def\Bl{\Bigl}
\def\Br{\Bigr}
\def\bl{\bigl}
\def\br{\bigr}

\usepackage{color}

\usepackage{amsmath,amssymb}

\usepackage{subcaption}
\begin{document}

\noindent {\it Astronomy Reports, 2023, Vol. , No. }
\bigskip\bigskip  \hrule\smallskip\hrule
\vspace{35mm}


\title{Schwinger Pair Production and Vacuum Birefringence around High Magnetized Neutron Stars\footnote{Paper presented at the Fifth Zeldovich meeting, an international conference in honor of Ya. B. Zeldovich held in Yerevan, Armenia on June 12--16, 2023. Published by the recommendation of the special editors: R. Ruffini, N. Sahakyan and G. V. Vereshchagin.}}

\author{\bf \firstname{Chul Min}~\surname{Kim}}%
\email{chulmin@gist.ac.kr}
\affiliation{Advanced Photonics Research Institute, Gwangju Institute of Science and Technology,  Gwangju 61005, Korea}
\affiliation{Center for Relativistic Laser Science, Institute for Basic Science, Gwangju 61005,  Korea}

\author{\bf \firstname{Sang Pyo}~\surname{Kim}}
\email{sangkim@kunsan.ac.kr}
\affiliation{Department of Physics, Kunsan National University, Gunsan 54150, Korea}
\affiliation{Asia Pacific Center for Theoretical Physics, Pohang 37673, Korea}
\affiliation{ELI Beamlines, ELI-ERIC, Doln\'{i} B\v{r}e\v{z}any, Czech Republic }

\begin{abstract}

\centerline{\footnotesize Received: ;$\;$
Revised: ;$\;$ Accepted: .}\bigskip\bigskip\bigskip
Highly magnetized neutron stars have magnetic fields of order of the critical field and can lead to measurable QED effects. We consider the Goldreich-Julian pulsar model with supercritical magnetic fields, induced subcritical electric fields, and a period of milliseconds. We then study the strong field physics, such as Schwinger pair production and the vacuum birefringence including the wrench effect, whose X-ray polarimetry will be observed in future space missions.
\end{abstract}

\maketitle

\section{Introduction}

Ultra-strong electromagnetic (EM) fields  abound in astrophysics. Neutron stars have strong magnetic fields that are stable for long duration and extend to astrophysical scale~(for review, see~\cite{Shapiro1983Black} and also~\cite{Meszaros1992High}), whose strength goes beyond the present laboratory achievements, $ I=1.1 \times 10^{23}\, {\rm W/cm^2}$~\cite{Yoon2021Realization}. Chandrasekhar and Fermi ~\cite{chandrasekhar1953problems} derived the upper bound for the magnetic fields from the virial theorem of magnetohydrostatic equilibrium that the magnetic energy of compact object can not exceed the gravitational binding energy, $B \leq 10^{18} \bl( M_{\rm N}/1.4 M_{\odot} \br) \bl(10\, {\rm km}/ R_{\rm N} \br)^2\, {\rm G}$. The dynamo can further amplify the magnetic field beyond  $10^{15}\, {\rm G}$, though the flux conservation during gravitational collapse of stars gives the magnetic fields about $10^{12}\, {\rm G}$ \cite{thompson1993neutron}. Recently, magnetars, highly magnetized neutron stars, have been observed \cite{vasisht1997discovery}, and those with supercritical magnetic fields and their astrophysical properties are listed in McGill Catalog~\cite{Olausen2014McGILL} and~\cite{Kaspi2017Magnetars}.

It has been known since the QED action was obtained by Heisenberg-Euler and Schwinger (HES) \cite{Heisenberg1936Folgerungen,Schwinger1951Gauge} that strong EM fields make the Dirac vacuum polarized as media, and that strong electric fields spontaneously create pairs of electrons and positrons. The HES action results in nonlinear electrodynamics that modifies the linear Maxwell theory. The criteria for necessity of nonlinear electrodynamics are  given as follows (in the Lorentz-Heaviside units with $c=\hbar=1$): (i) in plasma and matter systems, the magnetic or electric length $(1/\sqrt{eB}, 1/\sqrt{eE})$ for charge $e$ is comparable to or shorter than the characteristic length scale of physical systems, (ii)
in the Minkowski spacetime, the characteristic length is the Compton wavelength $(1/\sqrt{eB_c} = 1/\sqrt{eE_c}= 1/m)$ and QED loops contribute to the EM theory.
The critical electric and magnetic fields $E_c = m^2/e = B_c$ for a charge $e$ with mass $m$ have the electrostatic potential energy  over a Compton wavelength or the Landau levels equal to the rest mass: $E_c = 1.3 \times 10^{16}\, {\rm V/cm}$ and $B_c = 4.4 \times 10^{13}\, {\rm G}$ or the intensity $I_c = 2.3 \times 10^{29}\, {\rm W/cm^2}$. When electric or magnetic fields become comparable to or stronger than the critical field, the QED loop corrections should be included for accuracy. 

\begin{figure}
    \centering
    \begin{subfigure}{0.3\textwidth}
        \centering
        \includegraphics[width=\textwidth]{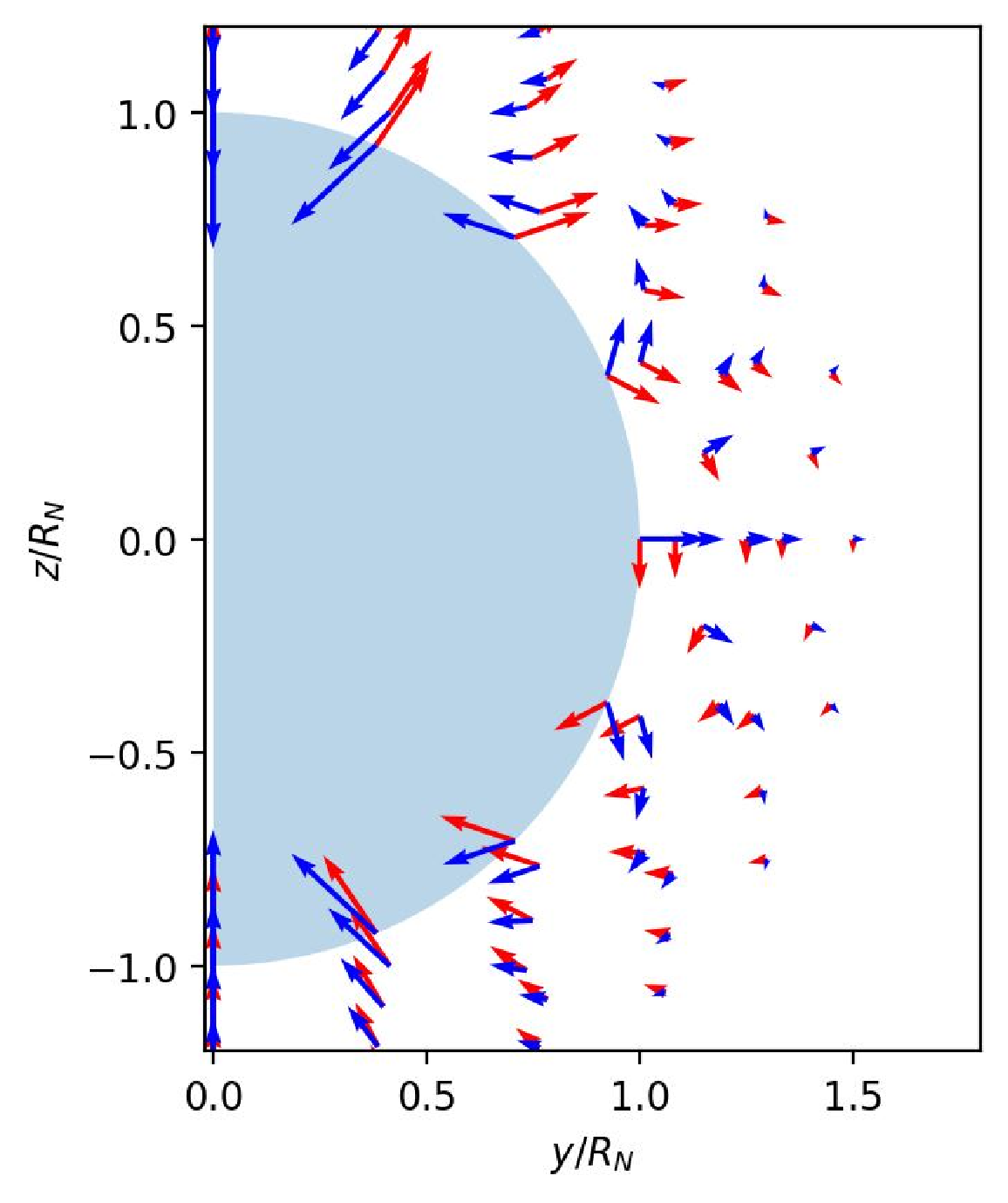}
        \caption{}
        \label{fig_EB}
    \end{subfigure}
    \hfill
    \begin{subfigure}{0.3\textwidth}
        \centering
        \includegraphics[width=\textwidth]{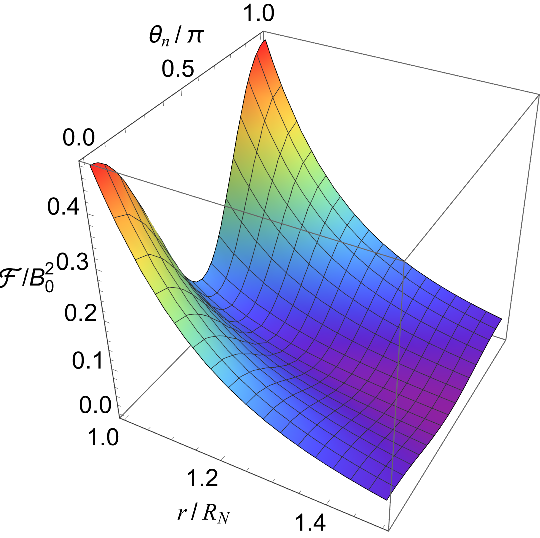}
        \caption{}
        \label{fig_F}
    \end{subfigure}
    \hfill
    \begin{subfigure}{0.3\textwidth}
        \centering
        \includegraphics[width=\textwidth]{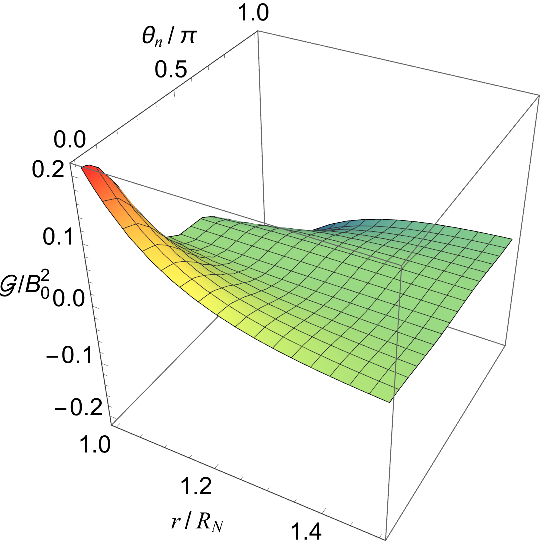}
        \caption{}
        \label{fig_G}
    \end{subfigure}
    \caption{ Electromagnetic fields around a pulsar. (a) EM fields of the Goldreich-Julian dipole model for pulsars. The red arrows denote the magnetic field (in units of $B_0$), and the blue ones denote the electric field (in units of $B_0 R_N \Omega /c$). $B_0$ refers to the field strength at the north pole, $R_N$ to the pulsar's radius, and $\Omega$ to the angular frequency of rotation; (b) $\mathcal{F}(r, \theta_n)$ and (c) $\mathcal{G}(r, \theta_n)$ in units of $B_0^2$ for $R_N \Omega /c = \pi/15/1000$, e.g., a neutron star with a radius of 10 km and a period of 1 s. }
\end{figure}

QED loop corrections result in nonlinear electrodynamics in which the vacuum acquires  field-dependent permittivity and permeability, in particular, the magneto-electric response. In the presence of strong electric fields, the vacuum decays due to spontaneous creation of electron-positron pairs. The nonlinear theory that respects the Lorentz- and gauge-invariance should be a functional of the Maxwell scalar ${\cal F}$ and the pseudo-scalar ${\cal G}$:
\b 
{\cal F} = \frac{1}{4} F_{\mu \nu} F^{\mu \nu} = \frac{1}{2}(\vec{B}^2-\vec{E}^2), \quad {\cal G} = \frac{1}{4} F_{\mu \nu} F^{* \mu \nu} = -\vec{E}\cdot\vec{B}.
\e
In this paper we will consider the dipole pulsar model by Goldreich and Julian (GJ)~\cite{Goldreich1969Pulsar}, as shown in Fig.~\ref{fig_EB}, whose ${\cal F}$ and ${\cal G}$ are
\b \label{GJ-FG}
{\cal F} &=& \f{B_0^2}{8} \Bl(\f{R_N}{r} \Br)^6 \Bl[ 1 + 3 \cos^2\theta_n - \Bl(\f{R_N}{r} \Br)^2 \Bl( \frac{R_N \Omega}{c} \Br)^2 \bl(1-2 \cos^2\theta_n + 5 \cos^4\theta_n \br) \Br], \nonumber\\
{\cal G} &=& B_0^2 \Bl(\f{R_N}{r} \Br)^{7} \Bl( \frac{R_N \Omega}{c} \Br) \cos^3\theta_n.
\e
Here, $R_N$ and $\Omega$  are the radius and angular frequency of a neutron star, and $\theta_n$ is the polar angle from the north pole (for simplicity, the magnetic polar axis assumed to be along the rotational axis, but see Ref.~\cite{Turolla2015Magnetars} for realistic configurations of magnetic fields). Note that the maximal surface velocity of neutron star (in units of $c$) can be written as
\b
\f{R_N \Omega}{c} = \f{\pi}{15} \times \Bl(\f{R_N}{R_{10}} \Br) \Bl( \f{P_{-3}}{P} \Br),
\e
where $R_{10} = 10\, \mathrm{km}$ and $P_{-3} = 10^{-3}\, \mathrm{sec}$ refers to millisecond neutron stars. Most neutron stars have $(R_N/R_{10})(P_{-3}/P) \leq 1$, so the maximal surface velocity does not violate the special relativity.
On the equatorial plane $(\theta_n = \pi/2)$, ${\cal G} = 0$, but $E^2(r)/B^2(r) = \bl({R_N}/{r} \br)^2 \bl({R_N \Omega}/{c} \br)^2$. In general, ${\cal G}/{\cal F} < 2 \bl({R_N}/{r} \br) \bl({R_N \Omega}/{c} \br) $, so we may assume ${\cal G} \ll {\cal F}$ away from the neutron stars.

The purpose of this paper is to further explore strong QED effects: the vacuum polarization and Schwinger pair production in supercritical magnetic fields and subcritical electric fields,   motivated by highly magnetized neutron stars, in particular, magnetars, which have recently been intensively studied in Refs.~\cite{Kim:2021kif,Kim:2022fkt}. To do so, we find a new analytical expression for HES action by keeping supercritical magnetic fields in  Refs.~\cite{Dittrich1976One,Dittrich1985Effective,Kim2011QED,Kim2019Quantum} but by expanding the part for the subcritical fields. We then study the vacuum birefringence for photons in EM fields of neutron stars. This study can be used in space missions that have recently been proposed to probe the QED regime, such as Compton Telescope \cite{wadiasingh2019magnetars} and the enhanced X-ray Timing and Polarimetry (eXTP) \cite{santangelo2019physics}. Physical processes in supercritical magnetic fields combined with induced electric fields differ from those in weak-field regime, for instance the curvature synchrotron radiation and QED cascade of pairs \cite{Meszaros1992High,Battesti2013Magnetic,Harding2006Physics}, not to mention the electron-positron pair production in astrophysics \cite{Ruffini2010Electronpositron}.


\section{QED Action in a Supercritical Magnetic Field and a Subcritical Electric Field}
\label{secII}

The vacuum polarization and Schwinger pair production in a constant EM field has been described by Heisenberg, Euler, and Schwinger's exact one-loop QED action
\b \label{HES}
{\mathcal L^{(1)}} (a,b) = - \f{1}{8 \pi^2} \int_{0}^{\infty} ds \f{e^{-m^2 s}}{s^3}
\Bl[(eas) \coth(eas)   (ebs) \cot(ebs) - 1 - \f{(es)^2}{3} (a^2 - b^2) \Br],
\e
where $a$ and $b$ are the gauge and Lorentz invariants
\b
a = \sqrt{\sqrt{{\cal F}^2 + {\cal G}^2} + {\cal F}}, \quad b = \sqrt{\sqrt{{\cal F}^2 + {\cal G}^2} - {\cal F}}.
\e
The QED action respects the CP invariance, ${\cal L}^{(1)} (\pm a, \pm b) = {\cal L}^{(1)} ( a, b)$, and the EM duality, ${\cal L}^{(1)} (a, b) = {\cal L}^{(1)} (ib, -ia)$~\cite{Kim2011QED,Kim2019Quantum}. 

Given an EM field configuration, $a$ and $b$ characterize in a gauge- and Lorentz-invariant way the magnetic field strength and the electric field strength, respectively. When ${\mathcal G} = 0 $, called electromagnetically wrenchless, a Lorentz frame can be found to remove the weaker between the electric field and the magnetic field: $a = B, b = 0$ when $\mathcal{F} >  0$ and $a = 0, b = E$ when $\mathcal{F}<0$. When ${\mathcal G} \neq 0 $, called electromagnetically wrenched, a Lorentz frame can be found to make the magnetic field parallel to the electric field: $a = B, b = E$. Since $a$ and $b$ determine the effective action and thereby the resultant phenomena,  we may regard $a$ the intrinsic magnetic field strength and $b$ the intrinsic electric field strength of a given field configuration, irrespective of  reference frames.  Hence, the configuration of  a supercritical magnetic field and a subcritical electric field is unequivocally represented by the condition  $ a > E_c > b $.

In a supercritical magnetic field with or without a subcritical electric field, (\ref{HES}) can be used for a numerical evaluation. Hence, it will be helpful to have an analytical expression for (\ref{HES}) in terms of known special functions. To do so, we expand $(ebs) \cot(ebs)$ in (\ref{HES}) using the power series 1.411 in \cite{Gradshtein2015Table}
\begin{equation}
x \cot x = 1 - \sum_{k = 1}^{\infty} \frac{2^{2k} |B_{2k}|}{(2k)!} x^{2k},
\end{equation}
where $B_{2k}$ are the Bernoulli numbers.
The leading term in the expansion of ${\cal L}^{(1)}(a,b)$,  i.e., ${\cal L}^{(1)}_{(0)}(a,0)$, was obtained in a closed form by Dittrich through dimensional regularization  \cite{Dittrich1976One,Dittrich1985Effective} and also by Kim and Lee using  the gamma function regularization of the in-out formalism \cite{Kim2019Quantum}:
\b
\label{L1a0_int}
{\cal L}^{(1)}_{(0)}(a,0) = \frac{m^{4}}{8\pi^{2}\bar{a}^{2}}\left[\zeta'(-1,\bar{a})-\frac{1}{12}+\frac{\bar{a}^{2}}{4}-\left(\frac{1}{12}-\frac{\bar{a}}{2}+\frac{\bar{a}^{2}}{2}\right)\ln\bar{a}\right].
\e
Here, $\bar{a}={m^2}/{2ea}$, $\zeta'(-1,\bar{a})=d\zeta(s,\bar{a})/ds|_{s=-1}$, and $\zeta(s,\bar{a})$ is the Hurwitz zeta function. The next leading term, proportional to $(eb)^2$, is
\b
\label{L1a1_int}
{\cal L}^{(1)}_{(1)}(a,b) = \frac{(eb)^2}{24\pi^{2}} \left[- \frac{1}{2 \bar{a}} +\ln{\bar{a}}- \psi (\bar{a}) \right], 
\e
where $\psi (z) = \Gamma'(z)/\Gamma(z)$ is the digamma function. The higher order terms are given by
\b
\label{L1a2_int}
{\cal L}^{(1)}_{(\geq 2)}(a,b) = \frac{(ea)(eb)}{4\pi^{2}} \sum_{k \geq 2}^{\infty} \frac{|B_{2k}|}{(2k) (2k-1)} \left[2 \Bl(\frac{b}{a} \Br)^{2k-1} \zeta \bl(2k-1, \bar{a} \br)-  \Br(\frac{2eb}{m^2} \Bl)^{2k-1} \right].
\e
Here, we have used the integral formula 3.551 in \cite{Gradshtein2015Table} and the definition of $\Gamma$-function:
\begin{equation}
\int_{0}^{\infty}\mathrm{d}x\ x^{\mu-1}e^{-\beta x}\coth x=\Gamma(\mu)\left[2^{1-\mu}\zeta\left(\mu,\frac{\beta}{2}\right)-\beta^{-\mu}\right]\quad\left[\mathrm{Re}(\mu)>1,\ \mathrm{Re}(\beta)>0\right].
\end{equation}
Note that the QED action was obtained in powers of ${\cal G}$  by Heyl and Hernquist~\cite{Heyl1997Analytic,Heyl1997Birefringence}.

\begin{figure}
	\includegraphics[width=0.6\textwidth]{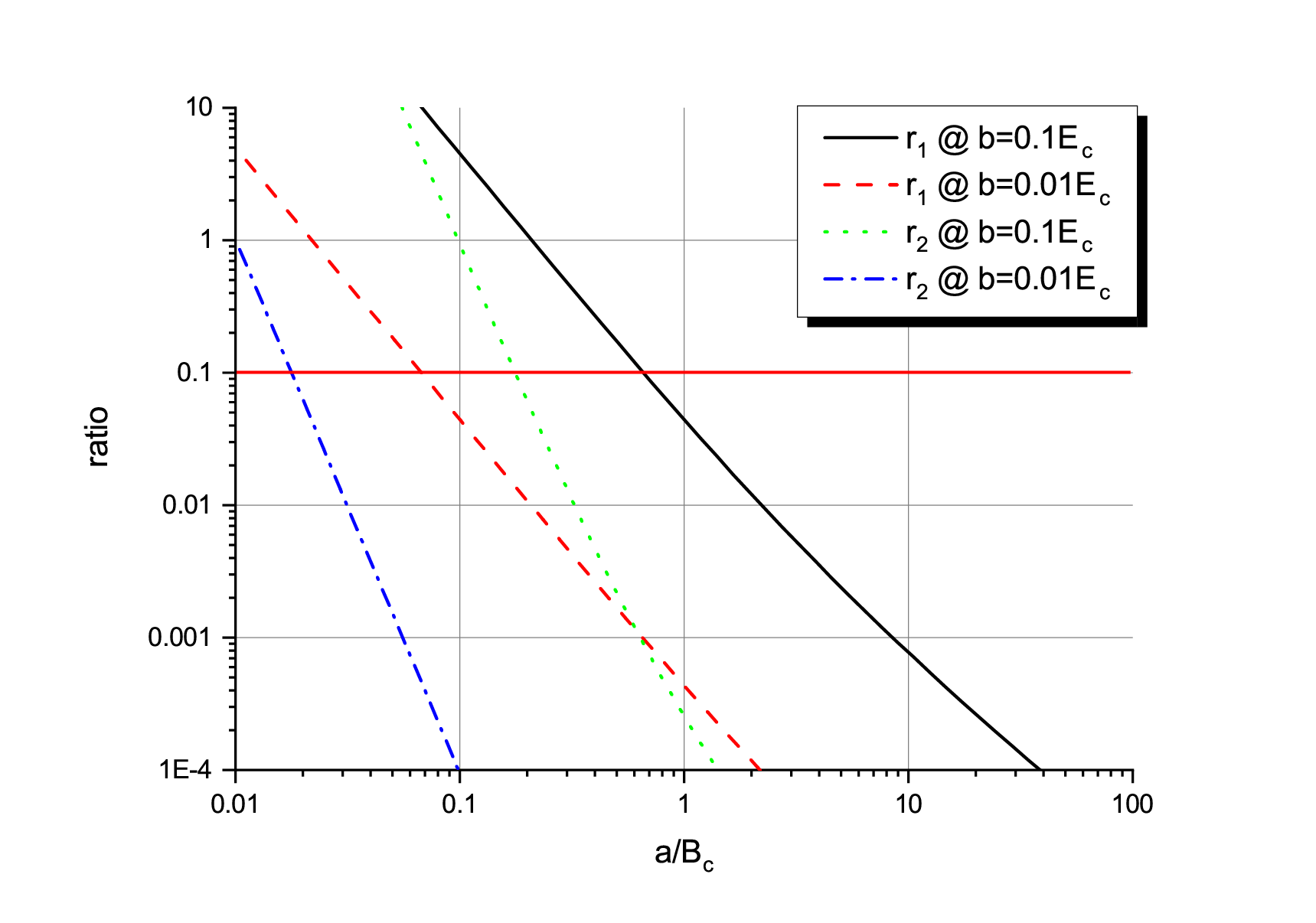}
	\caption{$ {\cal L}^{(1)}_{(1)}(a,b) / {\cal L}^{(1)}_{(0)}(a,b) $ (denoted by $r_1$) and $ {\cal L}^{(1)}_{(2)}(a,b) / {\cal L}^{(1)}_{(0)}(a,b)$ (denoted by $r_2$) for $ b=0.1E_c,~0.01E_c$. The ratio of 0.1, marked by the red thick horizontal line, may be considered as the threshold of small corrections.}
	\label{fig_ratio}
\end{figure}

The importance of the correction terms ${\cal L}^{(1)}_{(1)}$ and ${\cal L}^{(1)}_{(2)}$ with respect to ${\cal L}^{(1)}_{(0)}$ is exemplified in Fig.~\ref{fig_ratio} for $b =0.1E_c, ~ 0.01E_c$. We may take the ratio of 0.1 as the threshold of small corrections. When $b = 0.1E_c$, the first-order correction ${\cal L}^{(1)}_{(1)}$ is small compared to $ {\cal L}^{(1)}_{(0)}$ for $a \geq 0.66B_c$, whereas for $b = 0.01E_c$, it is small for $a \geq 0.068B_c$. These results empirically imply that $a \geq 7b$ is necessary to keep the correction terms less than 10\% of the leading term when $b\leq 0.1 E_c$, which is consistent with the assumption of a supercritical magnetic field and a subcritical electric field.  When ${\cal L}^{(1)}_{(1)}$ can be considered small, the correction ${\cal L}^{(1)}_{(2)}$ can be neglected as it is smaller than ${\cal L}^{(1)}_{(1)}$ by two orders, as shown in Fig.~\ref{fig_ratio}. Thus, including only the subleading correction ${\cal L}^{(1)}_{(1)}$ would be sufficient for an accurate calculation in the valid parameter range. At a higher value of $b$ closer to $E_c$, the constraint on $a$ becomes more demanding: $a > 8b$ is necessary to keep the correction small when $b=0.5E_c$. In such a case, however, the pair production due to the imaginary part of  ${\cal L}^{(1)}(a,b)$  becomes significant and thus should be taken into account (see the next section).

Remarkably, Ritus obtained the QED action in supercritical magnetic field up to the two-loop level~\cite{Ritus1976Lagrangian}
\b
{\cal L}^{(1)}+{\cal L}^{(2)} = \f{(eB)^2}{24 \pi^2} \Bl[ \Bl(1+\f{3 \alpha}{4 \pi}\Br)  \ln \Bl(\f{ eB}{\pi \gamma m^2} \Br) +
\f{6}{\pi^2} \zeta'(2) + \cdots \Br].
\e
Thus the two-loop action is smaller by a factor of $\alpha$ than the one-loop action, which implies the one-loop action can be used even for supercritical magnetic fields up to this small deviation. In reality, an electric field can hardly exceed the critical strength because of the electrical short induced by Schwinger pair production, near the critical field.
The dominant one-particle irreducible diagrams of all loops to the HES action by Karbstein~\cite{Karbstein2019All} replace $\alpha$ by $\alpha/(1 - (\alpha /3 \pi) \ln ( eB/m^2))$, which is illustrative in astrophysics applications.

\section{Schwinger Pair Production} \label{secIII}

The Schwinger pair production and vacuum decay originates from the poles of HES action (\ref{HES}).
The pair production rate per unit Compton four-volume is
\b
R (r, \theta_n; B_0/B_c) = \frac{\tilde{a} \tilde{b} m^4}{8 \pi^2} \coth \Bl(\pi \f{\tilde{a}}{\tilde{b}} \Br) e^{- \pi/\tilde{b}},
\e
where $\tilde{a} = a/B_c$ and $\tilde{b} = b/B_c$ are dimensionless parameters. The vacuum decay rate is given by the vacuum persistence amplitude
\b
2 \, \mathrm{Im} {\cal L}^{(1)} = \frac{\tilde{a} \tilde{b} m^4}{4 \pi^2} \sum_{n = 1}^{\infty} \coth \Bl(\pi \f{\tilde{a}}{\tilde{b}} \Br) e^{- \pi n /\tilde{b}},
\e
and in the case of $\tilde{b} = 0$ the sum can be expressed in terms of the polylogarithm function $(\tilde{a} m^2)^2/(4 \pi^2) L_2(e^{- \pi/\tilde{b}})$.

We consider only the EM field of the GJ model with (\ref{GJ-FG}), in which ${\cal F}$ $({\cal G})$ is symmetric (antisymmetric) with respect to the equatorial plane. Furthermore, the parameters $\tilde{a}$ and $\tilde{b}$ are proportional to $B_0/B_c$ and depend only the distance $r$ from the surface and the polar angle $\theta_n$. In this paper, we study a theoretical model for pulsars with the millisecond period and supercritical magnetic fields. The pair production is dominant in regions near the north and south poles because ${\cal G} < {\cal F}$ on the surface and ${\cal G} \ll {\cal F}$ away from the surface, so $\tilde{b} \simeq |{\cal G}/B_c^2| / \sqrt{2{\cal F}/B_c^2}$.

\begin{figure}[h]
\includegraphics[width=0.45\linewidth,height=0.35\textwidth]{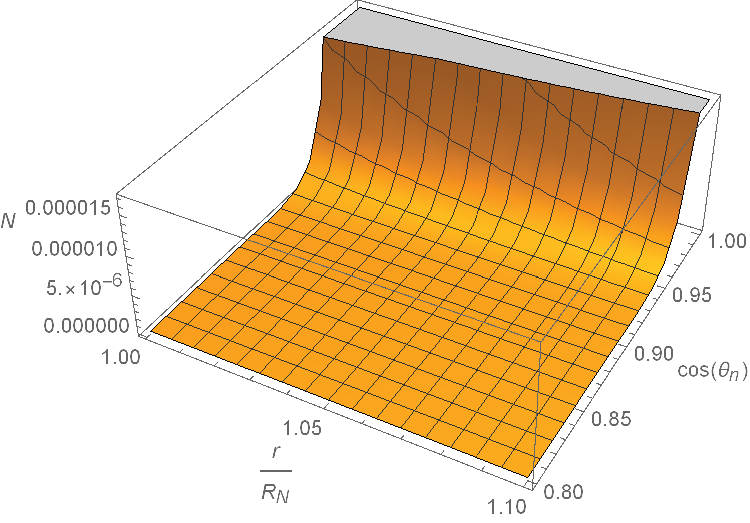}
\hfill
\includegraphics[width=0.45\linewidth,height=0.35\textwidth]{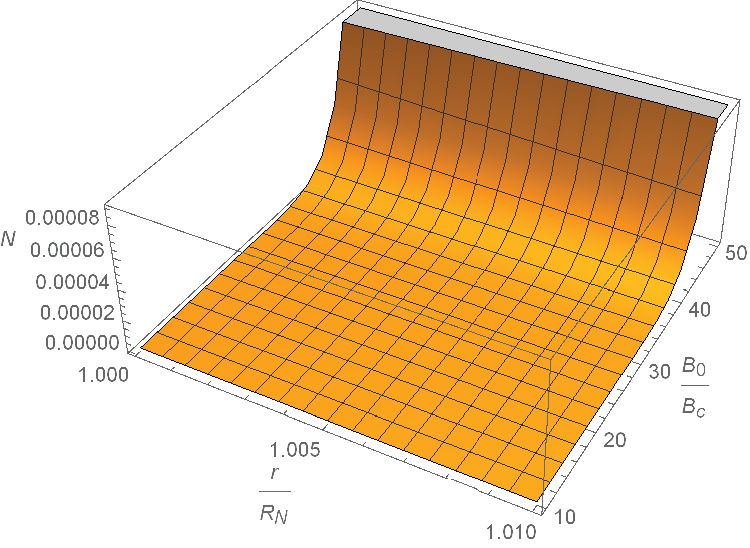}
\caption{The pair production rate per unit Compton four-volume against $r/R_N$ and $\cos \theta_n$ when $B_0/B_c = 50$ [left] and at poles against $r/R_N$ and $B_0/B_c$  [right].}
\label{N_r_cos}
\end{figure}

As shown in Figs.~\ref{fig_F} and \ref{fig_G}, ${\cal F}$ is maximal along polar axis and decreases away from the neutron star surface, and ${\cal G}$ vanishes on the equatorial plane and rapidly increases near the polar axis. This means that Schwinger pair production is significant close to the polar axis, as shown in Fig.~\ref{N_r_cos} (left panel). Because of the electric field direction at poles in Fig.~\ref{fig_EB}, near the north pole, positrons are emitted outward and electrons fall on the neutron star, and vice versa near the south pole. The emission forms ``jets'' of different charges along polar axis. The neutron star remains charge neutral because the pair production rate is symmetric with respect to the equatorial plane. The pair production rate increase by two order when the magnetic field strength increases from $B_0/B_c = 40$ to $B_0/B_c =50$ and the rate rapidly decreases away from the surface of neutron star as shown in Fig.~\ref{N_r_cos} (right panel).

The mean number of electrons or positrons must be multiplied by a huge factor from the astrophysical scale in units of the Compton four-volume $m^4 = 1/\lambda_{\rm C}^3 T_{\rm C} = 4.5 \times 10^{57}/ {\rm m^3\, sec} = 4.5 \times 10^{66} / {\rm km^3\, sec} $ at the north or south pole region. Roughly, one expects the number of electrons or positron on a pole per second ${\cal N}/{\mathrm{sec}} = 4.5 \times 10^{66} \times N$, so  ${\cal N}/{\mathrm{sec}} \sim  10^{20}$ even when $N = 10^{-44}$ which is produced by $B_0/ B_c = 6.08$.



\section{Vacuum birefringence for low-energy photons} \label{secIV}

The one-loop effective action $\mathcal{L}^{(1)}$ renders the vacuum under strong EM fields into an EM medium, of which total Lagrangian is given as
\begin{equation}
    \mathcal{L}=-\mathcal{F} + \mathcal{L}^{(1)}.
\end{equation}
Then, the polarization ($\vec{P}$) and magnetization ($\vec{M}$), a response of the medium, can be obtained by differentiating $\mathcal{L}$ with respect to $\vec{E}$ and $\vec{B}$ \cite{Greiner2009Quantum}:
\begin{equation}
	\vec{D}=\vec{E}+\vec{P}=\frac{\partial\mathcal{L}}{\partial\vec{E}},
	\quad
	\vec{H}=\vec{B}-\vec{M}=-\frac{\partial\mathcal{L}}{\partial\vec{B}},
 \label{DPHM}
\end{equation}
which, for $\mathcal{L}=\mathcal{L}(\mathcal{F},\mathcal{G})$, are reduced to
\begin{equation}
    \vec{D} = - \mathcal{L}_\mathcal{F} \vec{E} - \mathcal{L}_\mathcal{G} \vec{B}, \quad
    \vec{H} = - \mathcal{L}_\mathcal{F} \vec{B} + \mathcal{L}_\mathcal{G} \vec{E}.
     \label{eq_DH}
\end{equation}
The subscripts denote the differentiation with respect to the variable. These expressions imply that the vacuum is a nonlinear medium as $\mathcal{L}$ is a complicated function of $\vec{E}$ and $\vec{B}$.

We are interested in the propagation of a photon in such a nonlinear vacuum. When the photon energy is much smaller  than $m$, the photon cannot produce electron-positron pairs by interacting with the strong background field. Then, the photon can be treated as a weak classical perturbation to the background field. Splitting all the fields in (\ref{eq_DH}) into a strong background and a weak perturbation, $\vec{A}=\vec{A}_0 + \delta\vec{A}$, we can obtain the constitutive relation for the photon field:
\begin{equation}
    \delta \vec{D} = \tensor{\epsilon}_E \cdot \delta \vec{E} + \tensor{\epsilon}_B \cdot \delta \vec{B},
    \quad
    \delta \vec{H} = \tensor{\bar{\mu}}_B \cdot \delta \vec{B} + \tensor{\bar{\mu}}_E \cdot \delta \vec{E}.
    \label{eq_dDdH}
\end{equation}
It is assumed that the constitutive relations  (\ref{eq_DH}) hold separately for the background quantities and perturbation quantities. The tensor coefficients $\tensor{\epsilon}_E$, $\tensor{\epsilon}_B$, $\tensor{\bar{\mu}}_B$, and $\tensor{\bar{\mu}}_E$ are given in terms of $\vec{E}_0$, $\vec{B}_0$, and $\mathcal{L}_\mathcal{F}$,  $\mathcal{L}_\mathcal{G}$, $\mathcal{L}_\mathcal{FF}$, $\mathcal{L}_\mathcal{FG}$, $\mathcal{L}_\mathcal{GG}$, evaluated for $\vec{E}=\vec{E}_0$ and $\vec{B}=\vec{B}_0$ \cite{Kim31}. Note that $\tensor{\epsilon}_B$ and $\tensor{\bar{\mu}}_E$ give rise to the magneto-electric effect, which is rarely observed in usual optical media but has recently been observed in multiferroic materials \cite{Eerenstein2006Multiferroic,Fiebig2005Revival}.

Then, the Maxwell equations for the photon field are obtained as
\begin{equation}
    \frac{\partial\delta\vec{B}}{\partial t} = - \nabla \times \delta \vec{E}, \quad
    \frac{\partial\delta\vec{D}}{\partial t} = \nabla \times \delta \vec{H}.
\end{equation}
To find the photon propagation modes, we assume that all the perturbation quantities vary harmonically as $\exp(i\omega (n \hat{n} \cdot \vec{x}-t))$, where $\hat{n}$ is the direction of photon propagation, $\omega$ the photon frequency, and $n$ the refractive index to be determined. Then, the Maxwell equations are reduced to
\begin{equation}
    \delta \vec{B} = \vec{n} \times \delta \vec{E},\quad
    \delta \vec{D} = - \vec{n} \times \delta \vec{H},
    \label{eq_dBdD_harmonic}
\end{equation}
where $\vec{n}=n \hat{n}$.  Combining (\ref{eq_dDdH}) and (\ref{eq_dBdD_harmonic}) yields a homogeneous matrix-vector equation, whose solution is  the polarization vector $\delta \vec{E}$ and the refractive index $n$:
\begin{equation}
    \tensor{\Lambda} \cdot \delta \vec{E} =
    \tensor{\epsilon}_E \cdot \delta \vec{E}
    + \tensor{\epsilon}_B \cdot \vec{n} \times \delta \vec{E}
    + \vec{n} \times ( \tensor{\bar{\mu}}_B \cdot \vec{n} \times \delta \vec{E}
        + \tensor{\bar{\mu}}_E \cdot \delta \vec{E} )=0
\end{equation}
Albeit complicated, the equation can be solved exactly 
 through a systematic procedure  \cite{Kim31}.

For example, we consider a weak, purely magnetic background field, for which  the refractive indices and polarization vectors are obtained in simple forms. In such a case, $\mathcal{L}=-\mathcal{F} + \eta_1 \mathcal{F}^2 + \eta_2 \mathcal{G}^2$ , where $\eta_1/4=\eta_2/7=e^4/(360\pi^2 m^4)$ \cite{Adler1971Photon}.  Assuming  $\vec{B}_0 = B_0 \hat{z}$ and $\hat{n}=(\sin \theta,0,\cos \theta)$ without loss of generality, we obtain the modes as follows \cite{Kim31}:
\begin{equation}
\begin{split}
    n_1=\sqrt{\frac{1-\eta_{1}B_{0}^{2}}{1-\eta_{1}B_{0}^{2}-2\eta_{1}B_{0}^{2}\sin^{2}\theta}} \;\mathrm{with}\; \delta \vec{E}_1 & =(0,1,0) \\
    n_2=\sqrt{ \frac{1-\eta_{1}B_{0}^{2}+2\eta_{2}B_{0}^{2}}{1-\eta_{1}B_{0}^{2}+2\eta_{2}B_{0}^{2}\cos^{2}\theta}} \;\mathrm{with}\; \delta \vec{E}_2 & =\left( -\frac{(1-\eta_{1}B_{0}^{2}+2\eta_{2}B_{0}^{2})}{1-\eta_{1}B_{0}^{2}}\cot\theta,0,1\right).
\end{split}
\end{equation}
As $n_1 \neq n_2$, birefringence occurs. Note that $\hat{n}\cdot \delta\vec{E}_1 = \delta\vec{E}_1 \cdot \delta\vec{E}_2 =0$ but $\hat{n}\cdot\delta\vec{E}_2 \neq 0$.

When the background field is strong, $\mathcal{L}^{(1)}=\mathcal{L}^{(1)}_{(0)}+\mathcal{L}^{(1)}_{(1)}+\cdots$ in Sec.~\ref{secII} should be used to obtain the modes. For the same configuration of $\vec{B}_0$ and $\hat{n}$ as in the weak field case, the modes are given as follows \cite{Kim:2022fkt}:
\begin{equation}
\begin{split}
    n_1=\sqrt{\frac{1}{1+\mu \sin^2 \theta}} \;\mathrm{with}\; \delta \vec{E}_1 & =(0,1,0) \\
    n_2=\sqrt{\frac{1+\epsilon}{1+\epsilon \cos^2 \theta} } \;\mathrm{with}\; \delta \vec{E}_2 & =\left( -(1+\epsilon ) \cot \theta,0,1\right),
\end{split}
\end{equation}
where
\begin{equation}
    \mu = -\frac{B_0^2 \mathcal{L}^{(1)}_{\mathcal{FF}}}{1-\mathcal{L}^{(1)}_\mathcal{F}}, \quad
    \epsilon = \frac{B_0^2 \mathcal{L}^{(1)}_{\mathcal{GG}}}{1-\mathcal{L}^{(1)}_\mathcal{F}}.
\end{equation}
As the background field is purely magnetic, one may be tempted to drop the terms $\mathcal{L}^{(1)}_{\geq 1}$, which is null without an electric field, from the beginning. However, the fields in $\mathcal{L}$ are the combinations of the background field and the photon field, which definitely has an electric part due to its propagating nature. Thus, the expansion presented in Sec.~\ref{secII} is crucial even for the mode analysis with a purely magnetic background field. If an additional electric field is added to have  a component parallel to the strong magnetic field, the birefringence effect, $|n_1 - n_2 |$, increases, and the polarization vectors rotate \cite{Kim:2022fkt}.

The birefringence experienced by low-energy photons around highly magnetized neutron stars is a characteristic phenomenon of the nonlinear quantum vacuum. If observed in such environments, it will not only reveal the nature of the quantum vacuum under strong fields but also estimate the magnetic field  around compact astronomical objects. The formalism developed in this paper and a systematic mode analysis will provide an accurate framework for such goals.

\section{Conclusion}\label{secV}

Ultra-strong magnetic fields of neutron stars induce relatively strong electric fields due to their rapid rotation from milliseconds to seconds. Hence the strong QED plays an important role in understanding physics, such as vacuum polarization effects, electron-positron pair production, photon-photon interactions, interactions of charges in strong EM fields, particularly, the acceleration of charges and radiations and plasma QED etc. In this paper we have studied strong field QED physics in supercritical magnetic fields and induced subcritical electric fields, and particularly, two of the prominent features: the Schwinger pair production and the vacuum birefringence.

A theoretical model studied in this paper with a millisecond period and supercritical magnetic fields leads to the Schwinger pair production, the dominant emission mechanism of electron-positron pairs due to the astrophysical scale volume $\mathrm{km}^3$ in Compton units around the north and south poles. This model provides us with the most extreme environments for strong field QED. However, the confirmed magnetars with supercritical magnetic fields have the period of second or longer and millisecond neutron stars have magnetic fields weaker by order than the critical field, so the Schwinger effect is exponentially suppressed by a huge factor even for the astrophysical volume. On the other hand, the vacuum birefringence may give rise to observable polarization effects, which can be measured through X-ray polarimetry: space telescopes like Compton Telescope and eXTP are planned for that purpose. The precision observations of X-ray polarimetry will enable one to reconstruct EM field configurations of neutron stars and charged and/or magnetized black holes.


\begin{acknowledgments}
S.P.K. would like to thank Sergei Bulanov, Tae Moon Jung, Remo Ruffini, Rashid Shaisultanov, Gregory Vereshchagin for useful discussions. This work was supported by Institute for Basic Science (IBS) under IBS-R012-D1. The work of C.M.K. was also supported by Ultrashort Quantum Beam Facility operation program (140011) through APRI, GIST and GIST Research Institute (GRI) grant funded by GIST in 2023. The work of S.P.K. was in part supported by National Research Foundation of Korea (NRF) funded by the Ministry of Education (2019R1I1A3A01063183).
\end{acknowledgments}



\clearpage


\bibliography{refs_Zeldovich}
















\end{document}